\documentstyle[12pt]{article}
\begin{document}
\newcommand{\be}{\begin{equation}}
\newcommand{\ee}{\end{equation}}
\newcommand{\ba}{\begin{array}}
\newcommand{\ea}{\end{array}}
\newcommand{\gtrsim}{\; \raisebox{-1ex}
   {$ \stackrel{\textstyle >}{\sim}$} \;}
\newcommand{\lesssim}{\; \raisebox{-1ex}
   {$ \stackrel{\textstyle <}{\sim}$} \;}
\begin{titlepage}
\begin{flushright}
hep-th/9311136 \\
BI-TP 93/59 \\
\end{flushright}
\begin{center}
{\Large PARTICLES WITH DISTANCE DEPENDENT STATISTICS
AT LOW TEMPERATURES} \\
\vfill
{\large Stefan V.~Mashkevich}\footnote{~E-mail: mash@phys.unit.no,
gezin@gluk.apc.org}, {\large Gennady M.~Zinovjev}\footnote{~E-mail:
gezin@gluk.apc.org}\\
\bigskip
{\it Fakult\"at f\"ur Physik, Universit\"at Bielefeld \\
D-33501 Bielefeld, Germany \\ and \\
Institute for Theoretical Physics, Academy of Science of Ukraine \\
252143 Kiev, Ukraine}\/\footnote{~Permanent address}
\vfill
{\large Abstract}
\end{center}
\begin{quotation}
We consider a simplified model of particles with effectively
distance dependent statistics, that is particles coupled to
a gauge field the Lagrangian of which contains the Chern-Simons
term. We analyze the low-lying states of the two-particle
system and show that under certain conditions they can exhibit
negative compressibility, hinting on a possible \`a la van der
Vaals picture.
\end{quotation}
\vfill
\begin{flushleft}
November 1993
\end{flushleft}
\end{titlepage}

It is well known that particles coupled to a Chern-Simons
gauge field in (2+1) dimensions undergo an effective change of
statistics, i.e. become anyons \cite{Wil82,Aro85}.
A much more practical situation is that when the gauge
field Lagrangian is a sum of the Chern-Simons
term and of other term(s) of non-topological nature. In this
case there is always a characteristic scale of length and
the behaviour of the particles is different depending on
whether they are close to each other or far away on
this scale. In the two limiting cases they behave like
anyons but with different values of the statistical parameter.
By virtue of this it makes sense to speak about effective
``distance dependence of statistics'' for such particles
\cite{Mas93a}. This distance dependent statistics emerges in
various models, the simplest examples being
Maxwell-Chern-Simons electrodynamics \cite{Shi91,Mas92} and
the Dorey-Mavromatos model for high-$T_c$ superconductivity
\cite{Gor93}.

In a previous work \cite{Mas93a} the two-body problem for
particles with distance dependent statistics was considered,
special attention being paid to the high-temperature behaviour.
It was shown that in a semiclassical approximation a general
formula for the second virial coefficient of such particles
could be derived, which gives in the limiting cases the
result of Ref.\cite{Aro85} for anyons. In this letter we will
study the same system in the low-temperature regime, putting
the emphasis on the ground state, and show on a simplified model
that under certain conditions states with negative compressibility
can arise, hinting on a picture similar to the one with the
van der Vaals gas.

Thus, we consider a conserved current $j^{\mu}$ coupled to
a gauge field $A^{\mu},$ the total Lagrangian being
\be
{\cal L} = {\cal L}_0 + \frac{1}{2} \alpha
\epsilon^{\mu \nu \lambda} A_{\mu} \partial_{\nu} A_{\lambda}
- A_{\mu} j^{\mu} ,
\ee
where ${\cal L}_0$ depends on $A_{\mu}$ and its derivatives.
The corresponding field equations read
\be
Q^{\mu} + \alpha \epsilon^{\mu \nu \lambda} \partial_{\nu}
A_{\lambda} = j^{\mu} ,
\label{field_eqn}
\ee
where $Q^{\mu} = \partial_{\nu} \left( \partial {\cal L}_0
/ \partial (\partial_{\nu} A_{\mu}) \right) - \partial {\cal L}_0
/ \partial A_{\mu} .$ In accordance with these equations, a
static charge in the origin,
\be
j^{\mu} = e \delta_0^{\mu} \delta^2 (\vec{r}),
\label{stat_chrg}
\ee
gives rise, in general, to an electric
field (due to $Q^{\mu}$) as well as to a magnetic field
(due to $\epsilon^{\mu \nu \lambda} \partial_{\nu} A_{\lambda}$).
If ${\cal L}_0 $ does not depend explicitly upon time and
polar angle $\varphi,$ which is a natural assumption, then the
solution of (\ref{field_eqn}) with the right-hand side
(\ref{stat_chrg}) depends only on $r.$ In the Lorentz gauge
therefore one has $A_r \equiv 0.$ The temporal component $A_0$
corresponds to a usual charge-charge interaction, screened in the
presence of the Chern-Simons term and in any case irrelevant
to statistics. The angular component may always be written as
\be
A_{\varphi} (\vec{r}) = - \frac{\Delta(r)}{er} .
\label{A_phi}
\ee
The function $\Delta(r)$ possesses clear physical meaning:
$\Phi(r) = - \frac{2 \pi}{e}\Delta(r)$ is the magnetic  flux
created by the charge through the circle of radius $r.$
Therefore in any real model the above-mentioned
function is continuous and its limit values
\be
\Delta_0 = \Delta(0) \; , \; \Delta_{\infty} = \Delta(\infty)
\ee
are finite. If $\Delta(r) = const $
(this is the case if ${\cal L}_0 = 0$),
Eq. (\ref{A_phi})
describes the conventional Aharonov-Bohm potential, making the
particles effectively anyons. In the general case,
there is always a characteristic distance scale $d,$
which is essentially the size of the magnetic field cloud
created by the charge, so that $\Delta(r)$ may
be replaced by $\Delta_0 \, ( \Delta_{\infty} )$
if $r \ll d \; ( r \gg d).$

According to common quantum mechanical rules, the
Hamiltonian of the relative motion of two particles is
\be
\tilde{\cal H}_{2} = \frac{p^{2}_{r}}{m} +
\frac{\left[ p_{\varphi} + \Delta(r) \right]^{2}}{mr^{2}}
+ V(\vec{r}),
\ee
$V(\vec{r})$ being the mechanical interaction potential. If
the latter is central, then the angular part of the relative
wave function is separated as usually,
\be
\tilde{\Psi} = \exp(i \ell \varphi) \chi ,
\ee
and the levels $E^{\ell}_{n}$ ($\ell$ is the angular momentum,
$n$ the radial quantum number) are determined from the equation
for the radial part
\be
\tilde{\cal H}^{\ell}_{2} \chi = E^{\ell}_{n} \chi ,
\label{rel_eq}
\ee
where
\be
\tilde{\cal H}^{\ell}_{2} = \frac{p_{r}^{2}}{m}
+ \frac{\left[ \ell + \Delta(r) \right]^{2}}{mr^{2}}
+ V(r) ,
\label{Hamilt}
\ee
$p^{2}_{r} = -\partial^{2} / \partial r^{2} - (1/r)
\partial / \partial r,$ and $\ell$ is even (odd) for
bosons (fermions).

For arbitrary $\Delta(r)$ and $V(r)$ Eq.(\ref{rel_eq}),
naturally, cannot be solved exactly. The exact solution
is available, in particular, when the effective potential
energy in (\ref{Hamilt}),
\be
U^{\ell}(r) \equiv W^{\ell}(r) + V(r),
\ee
$W^{\ell}(r)$ being the centrifugal energy
(the second term in (\ref{Hamilt})), takes the oscillator form,
\be
U(r) = U_{0} + \frac{M^2}{mr^{2}} + \frac{m \Omega^{2} r^{2}}{4} .
\label{harmonic}
\ee
Then the ground state energy is
\be
E_{0} = U_{0} + (M+1)\Omega ,
\label{grst}
\ee
and the classically accessible region lies at both sides of
the point $r_{0} = \sqrt{2M/m \Omega}, $ which is the
minimum point of $U(r),$ and has the width
$\Delta r_{c} = 2/{\sqrt{m \Omega}}.$
With regard for this let us consider the case
of a harmonic external potential:
\be
V(r) = \frac{m \omega^{2} r^{2}}{4}.
\ee
If $\Delta(r) = const$ then $U^{\ell}(r)$ has just the form
(\ref{harmonic}) with $U_{0}=0 , M=|\ell + \Delta | ,
\Omega = \omega.$ One gets then
\be
E^{\ell}_{0} = \left( |\ell + \Delta | + 1 \right) \omega ,
\ee
and the ground state is that one for which the quantity
$|\ell + \Delta |$ is minimal (it is then certainly not
more than unity). This picture corresponds to anyons.
It is a good approximation when $\omega$ is such that
$\Delta(r)$ is
almost constant (more precisely, varies by a value
much less than unity) in the region where the probability
density for the distribution of $r$ differs considerably from
zero, that is, in a region of the width of the order
$\Delta r_{c}$ with the center in $r_{0}.$ This is
deliberately the case if all this region lies either at
$r \ll d$ or at $r \gg d$ (that is, if either $|\Delta(r)
- \Delta_{0}| \ll 1$ or $|\Delta(r) - \Delta_{\infty}|
\ll 1.$) In the intermediate situation, when $1/ \sqrt{m
\omega} \sim d ,$ the variation of $\Delta(r)$ in the
classically accessible region is a value of the order
$|\Delta'(d)| \! \cdot \! d.$ Since $\Delta(r)$ changes from
$\Delta_{0}$ to $\Delta_{\infty}$ within a region of the
width again of the order $d,$ the value under consideration
is, generally speaking, of the order $|\Delta_{\infty}
- \Delta_{0}|.$ Smallness of this compared to unity would
mean that $\Delta(r)$ is almost constant everywhere. So far
there is the simple conclusion: The particles behave like
anyons when they are at a distance either much more or much
less than the size of the cloud; at intermediate distances
such behaviour takes place only in the trivial situation when
the statistics weakly depends on distance.

Now we proceed to consider the general case. In order to carry
on the analytical treatment, we introduce the simplified model
(called ``extended anyons'' in Ref.\cite{Tru92}) with
\be
\Delta(r) = \left\{
	    \begin{array}{ccc}
	       D \frac{r^{2}}{d^{2}} & , & 0 \le r \le d , \\ \\
	       D & , & r \ge d .
	    \end{array}  \right.
\ee
The physical meaning is obvious: With each particle a magnetic field
is associated, uniformly distributed within a circle of radius
$d$ the center of which is the position of the particle. One gets
ideal anyons in the limit $d \rightarrow 0.$ The function $U^{\ell}(r)$
for $0 \le r \le d$ has the form (\ref{harmonic}) with
\be
U_{0} = G\ell \; , \; M = |\ell| \; , \;
\Omega = \sqrt{G^{2}+\omega^{2}} \; ,
\label{r_le_d}
\ee
where
\be
G = 2D \xi \; , \; \xi = \frac{1}{md^{2}} \; ,
\ee
and still the same form for $r \ge d$ but with
\be
U_{0} = 0 \; , \; M = |\ell +D| \; , \; \Omega = \omega \; .
\label{r_ge_d}
\ee
Consider the behaviour of the ground state with the following
assumptions: 1) the particles themselves are bosons; 2) $D \gg 1;$
3) $\delta \equiv D \bmod 2 < 1$ (the latter is exclusively for
simplifying the formulas). The derivative of the centrifugal
energy $W^{\ell}(r)$ is
\be
{W^{\ell}}'(r) = \frac{2[\ell + \Delta(r)]}{mr^{2}}
\left\{ \Delta'(r) - \frac{\ell +\Delta(r)}{r} \right\}
\ee
and, naturally, is discontinuous in the point $r=d.$ For
$\omega \ll \xi$ one has anyons with statistical parameter
$\delta$ ; the ground state corresponds to $\ell = - {\cal D}$
where ${\cal D} = D - \delta$ is the entire part of $D,$ and
has the energy
\be
E_{0} = (1+\delta)\omega .
\ee
The external parameter of the system is $1/ \omega$
(for the oscillator, $1 / m\omega$ is the area occupied
by the system, up to a constant factor),
and the corresponding external force (pressure) is
\be
p = \omega^{2} \frac{\partial E_{0}}{\partial \omega} ;
\ee
in the case at hand
\be
p = (1+\delta) \omega^{2} .
\ee
For $\omega \ll G,$ when $D(\Delta r_{c})^{2} / d^{2} \ll 1,$
one has bosons, the ground state of which is that with $\ell = 0,$
its energy is
\be
E_{0} =\omega ,
\ee
and the corresponding pressure is
\be
p =\omega^{2}.
\ee
Now let us take the intermediate case. At $-{\cal D} \le \ell \le 0$
the function $U^{\ell}(r)$ has a minimum in the region
$0 \le r \le d, $ and the energy of the corresponding state may
be found upon substituting (\ref{r_le_d}) into (\ref{grst}):
\be
E^{\ell}_{0} = G\ell + \{ |\ell| + 1\} \sqrt{G^{2}+\omega^{2}} .
\label{E0l1}
\ee
But for this state to exist really it is
necessary that the whole of the correspondent classically
accessible region be situated at $r \ll d.$ This is the case
for $|\ell| \ll {\cal D},$ and in particular for $\ell = 0,$
when the expression (\ref{E0l1}) reaches its minimum,
\be
E_{0}' = \sqrt{G^{2}+\omega^{2}},
\label{E0pr}
\ee
but not for $|\ell| \simeq {\cal D}.$ With this latter
option one has
\be
\begin{array}{rcl}
W^{\ell}(d) & = & \xi (\ell + D)^{2} , \\
{W^{\ell}}' (d-0) & \simeq & 4 D \xi (\ell + D) / d , \\
{W^{\ell}}' (d+0) & = & - 2 \xi (\ell + D)^{2} / d .
\end{array}
\ee
The question is when $U^{\ell}(r)$ may have a second minimum
at $r \gtrsim d,$ which a state with energy less than $E_{0}'$
would correspond to. If $\omega$ is small enough so that
$m \omega^{2}d / 2 < \left| {W^{\ell}}'(d+0)\right| ,$
then the minimum exists at $r > d,$ and the corresponding
energy is $(|\ell+D|+1)\omega;$ this is the above-mentioned
case of anyons at large distances. If $\ell+D > 0,$ then
$ {W^{\ell}}'(d-0) > 0$ and there can be no other minima
except for the two mentioned ones. But if $\ell+D < 0$ and
$ \left| {W^{\ell}}'(d+0) \right| < m\omega^{2}d / 2
< \left| {W^{\ell}}'(d-0) \right| ,$ then a minimum at
$r=d$ exists, which is a cusp point at the same time.
The energy of the corresponding state is minimal for
minimal $|\ell+D|$ and under the condition
$\xi^{2} \ll \omega^{2} \ll G\xi$ it equals
\be
E_{0}'' = \frac{m\omega^{2}d^{2}}{4} + \chi(\omega)
= \frac{\omega^{2}}{4\xi} + \chi(\omega) ,
\ee
where $\chi(\omega)$ is the heigth of the level above the
minimal value of $U^{\ell},$ which, as is easy to estimate
using the uncertainty relation, is proportional to
$\omega^{4/3}\xi^{-1/3}$ and because of $\omega \gg \xi$
is a small correction. Since $\omega^{2} \ll G\xi,$
the ground state energy is $E_{0}''$, because
$E_{0}'' < E_{0}'.$ The pressure is
\be
p \simeq \frac{\omega^{3}}{2\xi} .
\ee
Choosing, for example, $\omega^{2}_{1} \sim G\xi \cdot
\left( \xi \over G \right)^{1/4},$ one has
$p_{1} \sim G\xi \cdot \left( G \over \xi \right)^{1/8}
\sim G \xi \cdot D^{1/8}.$ Now let $G \xi \ll \omega^{2}
\ll G^{2};$ then the ground state energy will be
$E_{0}' \sim G,$ and the pressure
\be
p \simeq \frac{\omega^{3}}{\sqrt{G^{2}+\omega^{2}}}
\simeq \frac{\omega^{3}}{G}.
\ee
For $\omega^{2}_{2} \sim G \xi \cdot D^{1/4}$ one will
have $p_{2} \sim G \xi \cdot D^{-1/8}.$ Thus,
$\omega^{2}_{2} / \omega^{2}_{1} = D^{1/2}$ and at the
same time $p_{2} / p_{1} = D^{-1/4}.$
A situation arises which is similar to the one with
the van der Vaals gas: When the area is being decreased
(the frequency increased) the pressure decreases in
a certain region. When $\omega^{2} \ll G\xi, $ the
ground state is $E_{0}''$ and the mean distance between
the particles is of the order of
$d = 1/ \sqrt{m\xi} \gg 1/ \sqrt{m\omega},$ and
when $\omega^{2} \gg G\xi, $ the ground state is $E_{0}'$
and the distance is of the order of
$ 1/ \sqrt{mG} \gg 1/ \sqrt{m\omega}.$
When passing from the first regime to the second one
the particles undergo a ``condensation''.
In other words, the interaction of the particles
takes the character of a strong attraction. Apparently
it has to lead to the same effect of ``condensation''
for the $N$-particle system, which demands a special
consideration.

We would like to express our sincere gratitude to
J.~Engels, F.~Karsch, B.~Petersson and H.~Satz for
interesting discussions. This work was supported in
part by NATO Linkage Grant No.930224.

\end{document}